\documentstyle[preprint,aps]{revtex}
\newcommand{\bc}{\begin{center}}
\newcommand{\ec}{\end{center}}
\newcommand{\beqn}{\begin{equation}}
\newcommand{\eeqn}{\end{equation}}
\newcommand{\barr}{\begin{eqnarray}}
\newcommand{\earr}{\end{eqnarray}}

\def\del{\partial}
\def\etal {{\it et al}. }
\def\eg {{\it e.g}. }
\def\ie {{\it i.e}. }

\def\tr {\mbox{tr}}

\def\PL #1 #2 #3 {Phys. Lett.~{\bf#1} (#2) #3}
\def\NP #1 #2 #3 {Nucl. Phys.~{\bf#1} (#2) #3}
\def\NPP #1 #2 #3 {Nucl. Phys.~{\bf B} (Proc.~Suppl.)~{\bf#1} (#2) #3}
\def\ZP #1 #2 #3 {Z.~Phys.~{\bf#1} (#2) #3}
\def\PR #1 #2 #3 {Phys. Rev.~{\bf#1} (#2) #3}
\def\PP #1 #2 #3 {Phys. Rep.~{\bf#1} (#2) #3}
\def\PRL #1 #2 #3 {Phys. Rev.~Lett.~{\bf#1} (#2) #3}
\def\PTP #1 #2 #3 {Prog. Theor.~Phys.~{\bf#1} (#2) #3}
\def\MPL #1 #2 #3 {Mod. Phys.~Lett.~{\bf#1} (#2) #3}
\def\IJM #1 #2 #3 {Int. J.~Mod.~Phys.~{\bf#1} (#2) #3}
\input epsf
\begin{document}
\draft
\preprint{Submitted to Phys. Lett. {\bf B}}
\title{ 
Lattice Study of ${\rm U_{A}}(1)$ Anomaly: \\
The Role of QCD-Monopoles}
\author{Shoichi~SASAKI$^{\;a,\;b)}$\footnote[3]{E-mail address
:~ssasaki@bnl.gov}
 and Osamu~MIYAMURA$^{\;c)}$}
\address{a) Yukawa Institute for Theoretical Physics, Kyoto 
University, Kyoto 606-8502, Japan}
\address{b) RIKEN BNL Research Center, Brookhaven National 
Laboratory, Upton, NY 11973, USA}
\address{c) Department of Physics, Hiroshima University, 
Higashi-Hiroshima 739-0046, Japan\\
}
\date{September, 1998}
\maketitle
\renewcommand{\thefootnote}{\fnsymbol{footnote}}
\newlength{\minitwocolumn}
\setlength{\minitwocolumn}{7.0cm}
\setlength{\columnsep}{0.8cm}
\renewcommand{\thefigure}{\arabic{figure}} 
\newcommand{\fcaption}[1]{\refstepcounter{figure} Fig.~\thefigure  #1}
\begin{abstract}
We investigate the role of QCD-monopoles for the ${\rm U}_{\rm A}(1)$ 
anomaly in the maximally abelian gauge within the SU(2) lattice gauge theory.
The existence of the strong correlation between
instantons and QCD-monopoles in the abelian gauge
was already shown by both analytic
and numerical works including the Monte Carlo simulation.
Their interrelation brings us a conjecture that the presence of QCD-monopoles 
plays a considerable role on the ${\rm U}_{\rm A}(1)$ anomaly.
We find an evidence for our conjecture on a determination of
the fermionic zero modes of the Dirac operator in both the ``monopole removed''
gauge configuration and the ``photon removed'' gauge configuration.
\end{abstract}

\vspace{0.5cm}

\vfill\eject
\section{Introduction}

\indent
In non-abelian ${\rm SU}(N_{c})$ gauge theory,
any finite-action configuration is classified by the topological charge $Q$,
which is associated with the 
homotopy group $\pi_{3}({\rm SU}(N_{c}))=Z_{\infty}$, 
in the four-dimensional Euclidean space ${\bf R}^4$ \cite{Raja}. 
Instanton configurations are 
well known as classical and non-trivial 
gauge configurations, which satisfy the condition that the action is 
minimized in each sector $Q$ \cite{Raja}. 
We are reminded that the topological charge is equal to
the index of the massless Dirac operator 
$D \kern -2.2mm {/}$:
%
%
\beqn
Q = {\rm Index}\left[ D \kern -2.2mm {/} \right] \equiv n_{+} - 
n_{-}\;\;,
\eeqn
where $n_{+}(n_{-})$ is the number of zero-modes with $+(-)$ chirality.
This simple relation is known as the Atiyah-Singer index theorem 
\cite{Raja}.
The existence of the fermionic zero modes has the consequence 
that the global ${\rm U}_{\rm A}(1)$ symmetry is regarded as explicitly broken
at the quantum level owing to the ${\rm U}_{\rm A}(1)$ anomaly 
\cite{Hooft1}. 
Thus, such a topological feature 
implies that instantons are important topological objects in QCD
related to the resolution of the ${\rm U}_{\rm A}(1)$ problem 
\cite{Hooft1}.

Recently, some interesting results turn our attention to 
the non-trivial relation between instantons and magnetic monopoles 
\cite{Brower,Suganuma1,Miyamura1,Markum,Bornyakov,Suganuma2,Teper,Fukushima}.
As for the appearance of magnetic monopoles (QCD-monopoles) 
in ${\rm SU}(N_{c})$ gauge theory, 't Hooft proposed a
stimulating idea of the abelian gauge fixing \cite{Hooft2}.
Such a partial gauge fixing is defined by the gauge transformation in
the coset space of the gauge group to fix the gauge degrees
of freedom up to the maximally abelian subgroup; ${\rm U}(1)^{N_{c}-1}$.
In the abelian gauge, point-like singularities in the 
three-dimensional space ${\bf R}^3$ under the maximally abelian subgroup
can be identified as magnetic monopoles related
to the homotopy group $\pi_{2}({\rm SU}(N_{c})/{\rm U}(1)^{N_{c}-1})
=Z^{N_{c}-1}_{\infty}$ \cite{Hooft2}. 
The lattice simulations show that QCD-monopoles play a crucial role 
on color confinement in the QCD vacuum, which can be characterized by 
their condensation (see, \eg a recent review article 
\cite{Polikarpov}).

It is generally believed that instantons and QCD-monopoles are 
hardly thought to be associated with each other because these topological objects are 
originated from different non-trivial homotopy groups.
However, the recent analytical works 
have demonstrated the QCD-monopole as a classically stable solution 
in the background fields of the instanton configuration using
the abelian gauge fixing 
\cite{Brower,Suganuma1}. 
Furthermore, the several lattice simulations have
shown the strong correlation between 
instantons and QCD-monopoles in the highly quantum vacuum 
\cite{Suganuma1,Miyamura1,Markum,Bornyakov,Suganuma2}
as well as the semi-classical vacuum \cite{Suganuma2,Teper,Fukushima}.
Their interrelation brings us a conjecture that the presence of QCD-monopoles 
plays a considerable role on the ${\rm U}_{\rm A}(1)$ anomaly.

The main purpose of this paper is to find an evidence for such a conjecture
through the Monte Carlo simulation within the SU(2) 
lattice gauge theory. We examine the low-lying eigenvalue spectra of the Dirac 
operator to study the existence of the 
fermionic zero modes in both the ``monopole removed'' gauge configuration
and the ``photon removed'' gauge configuration, which are defined 
subsequently.
Although the Atiyah-Singer index theorem is not well-defined 
on the lattice in the strict sense, it might be approximately 
inherited at the finite lattice spacing.
We then expect that the relation between instantons and QCD-monopoles 
can be reexamined through an investigation of the
${\rm U}_{\rm A}(1)$ anomaly. Here, it is worth mentioning 
that whereas the measurement of the topological charge on the lattice 
usually needs some method to smooth Monte Carlo configurations,
the fermionic zero modes can be determined 
without any cooling method.

\section{Maximally abelian projection}

The Maximally Abelian (MA) gauge fixing \cite{Hooft2} 
was advocated by 't Hooft in order to define magnetic monopoles 
in the renormalizable and the Lorentz 
invariant way in the continuum:
$(\del_{\mu}\pm igA_{\mu}^{3})A_{\mu}^{\pm}=0$
where $A_{\mu}^{\pm}=A_{\mu}^{1}\pm iA_{\mu}^{2}$.
In the lattice formulation \cite{Kronfeld}, this gauge fixing 
corresponds to the maximization of the gauge dependent variable $R$;
%
%
\beqn
R[\Omega]=\sum_{n,\;\mu}\tr\left\{
\sigma_{3} U^{\Omega}_{\mu}(n) \sigma_{3} U^{\Omega \; \dag}_{\mu}(n) 
\right\}\;\;,
\eeqn
through the gauge transformation; 
$U_{\mu}(n) \rightarrow U^{\Omega}_{\mu}
(n)=\Omega(n)U_{\mu}(n)\Omega^{\dag}(n+{\hat \mu})$ where 
$U_{\mu}$ denotes the SU(2) link variable.
Once the gauge transformation is carried out to satisfy the above 
condition, the resulting SU(2) link variable $U^{\Omega}_{\mu}$ is factorized
into an 
abelian link variable $u_{\mu}(n)\equiv
\exp \{ i\sigma_{3}\theta_{\mu}(n) \}$ 
and an adjoint ``matter'' field $M_{\mu}$ 
\cite{Kronfeld} as
%
%
\beqn
U^{\Omega}_{\mu}(n)=M_{\mu}(n)\cdot u_{\mu}(n) \;\;,
\eeqn
where
%
%
\beqn
M_{\mu}(n) \equiv \left(
\begin{array}{cc}
\sqrt{1-|\xi_{\mu}(n)|^2} & \xi_{\mu}(n)\\
-\xi^{\ast}_{\mu}(n) & \sqrt{1-|\xi_{\mu}(n)|^2}
\end{array}
\right) \;\;.
\label{off-diag}
\eeqn
Performing the residual U(1) gauge transformation,
$\theta_{\mu}$ and $\xi_{\mu}$ respectively transform like an abelian gauge 
field and a charged ``matter'' field \cite{Kronfeld}.

Our next task is to look for the magnetic monopole in terms of
the ${\rm U}(1)$ variables. We consider the product of 
${\rm U}(1)$ link variables
around an elementary plaquette \cite{Kronfeld},
%
%
\beqn
u_{\mu \nu}(n)=
u_{\mu}(n) u_{\nu}(n+{\hat \mu})
u^{\dag}_{\mu}(n+{\hat \nu}) u^{\dag}_{\nu}(n)=e^{i\sigma_{3}
\Theta_{\mu \nu}(n)} \;\;.
\eeqn
It is worth mentioning that the ${\rm U}(1)$ plaquette $u_{\mu \nu}$
is a multiple valued 
function as the U(1) plaquette angle \cite{Kronfeld}; 
$\Theta_{\mu \nu}(n)\equiv \del_{\mu}\theta_{\nu}(n) - 
\del_{\nu}\theta_{\mu}(n) \in[-4\pi, 4\pi)$ where $\del$ denotes the
nearest-neighbor forward difference operator.
We then divide $\Theta_{\mu \nu}$ into 
two parts as
%
%
\beqn
\Theta_{\mu \nu}(n)={\bar \Theta}_{\mu \nu}(n)+2\pi n_{\mu 
\nu}(n) \;\;,
\label{Eq:Decomp}
\eeqn
where ${\bar \Theta}_{\mu \nu}$ 
is defined in the principal domain $[-\pi, \pi)$, 
which corresponds to the U(1) field strength in the continuum limit.
The integer-valued $n_{\mu \nu}$ is restricted as
$n_{\mu \nu}=0,\pm1,\pm2$ \cite{DeGrand}.
In terms of $\bar \Theta_{\mu \nu}$, 
the electric current $j_{\mu}$ and 
the magnetic current $k_{\mu}$ are defined as
%
%
\barr
j_{\mu}(n)&=&\del^{\prime}_{\nu}{\bar \Theta}_{\mu \nu}(n) \;\;,
\label{Eq:Ecurrent}\\
k_{\mu}(n)&=&\frac{1}{4\pi} \varepsilon_{\mu \nu \rho 
\sigma}\del_{\nu}{\bar \Theta}_{\rho \sigma}(n+{\hat \mu}) \;\;,
\label{Eq:Mcurrent}
\earr
where $\del^{\prime}$ denotes the nearest-neighbor backward difference
operator \cite{DeGrand}. 
Because of the Bianchi identity on the U(1) plaquette angle;
$\varepsilon_{\mu \nu \rho \sigma}\del_{\nu}
\Theta_{\rho \sigma}=0$, the magnetic current is rewritten as
%
%
\beqn
k_{\mu}(n)=-\frac{1}{2} \varepsilon_{\mu \nu \rho \sigma}\del_{\nu}n_{\rho 
\sigma}(n+{\hat \mu}) \;\;.
\label{monopole}
\eeqn
Eq.(\ref{monopole}) implies that the magnetic current,
which carries some integer values,
can be identified to the monopole trajectory
located on the boundary of the Dirac sheet; 
${}^{*}n_{\mu \nu}\equiv \frac{1}{2}\varepsilon_{\mu \nu \rho 
\sigma}n_{\rho \sigma}$ 
\cite{DeGrand}. The magnetic current is 
topologically conserved; $\del^{\prime}_{\mu}k_{\mu}(n)=0$ so that
the monopole trajectory forms a closed loop in the four-dimensional 
Euclidean space.

\section{Photon and monopole contribution}

Next, we aim to decompose the abelian gauge field into the
regular (photon) part and the singular (monopole) part \cite{Smit,Suzuki}.
We first perform the Hodge decomposition
on the U(1) field strength ${\bar \Theta}_{\mu \nu}$ \cite{Smit} as
%
%
\beqn
{\bar \Theta}_{\mu \nu}(n)=\del_\mu \theta^{\prime}_{\nu}(n)
-\del_\nu \theta^{\prime}_{\mu}(n) 
+ \varepsilon_{\mu \nu \rho \sigma}
\del^{\prime}_{\rho} B_{\sigma}(n)
\eeqn
with the dual gauge field $B_{\mu}$ satisfying the following equation \cite{Smit}:
%
%
\beqn
\left( \del^2 \delta_{\mu \nu} - \del^{\prime}_{\mu} \del_{\nu}
\right) B_{\nu}(n)=-2\pi k_{\mu}(n-{\hat \mu}) \;\;,
\eeqn
where $\del^2 = \del^{\prime}_{\mu} \del_{\mu}$.
The Gaussian fluctuation $\theta^{\prime}_{\mu}$ contributes
only the electric current $j_{\mu}$ and not the magnetic current $k_{\mu}$
so that it just corresponds to the regular (photon) 
part of the abelian gauge field. As a result, the singular (monopole) part of the abelian 
gauge field can be identified by subtracting the Gaussian fluctuation 
from the abelian gauge field \cite{Smit,Suzuki}.

In the Landau gauge, the definite identification of the Gaussian fluctuation, 
\ie the regular (photon) part, is given
by convolution of the U(1) field strength ${\bar \Theta}_{\mu \nu}$ with
the lattice Coulomb propagator $G(n-m)$ \cite{Smit,Suzuki} as
%
%
\beqn
\theta^{\rm Ph}_{\mu}(n) \equiv - \sum_{m,\;\lambda}G(n-m)
\del^{\prime}_{\lambda}{\bar \Theta}_{\lambda \mu}(m) \;\;.
\eeqn
Here the lattice Coulomb propagator satisfies the equation; 
$\del^{2}G(n-m)=-\delta_{n,\;m}$.
Immediately, we can obtain the singular (monopole) part from the 
following definition \cite{Smit,Suzuki}:  
%
%
\barr
\theta^{\rm Mo}_{\mu}(n) 
&\equiv& \theta^{L}_{\mu}(n) - \theta^{\rm Ph}_{\mu}(n) \nonumber \\
&=& -2\pi\sum_{m,\;\lambda}G(n-m)\del^{\prime}_{\lambda}
n_{\lambda \mu}(m) \;\;.
\earr
Here $\theta^{L}_{\mu}$ denotes the abelian gauge field in the Landau gauge
where $\del_{\mu}\theta^{L}_{\mu}(n)=0$. The singular part 
$\theta^{\rm Mo}_{\mu}$ actually carries the same
amount of the magnetic current 
as the original abelian gauge field in the infinite volume limit \cite{Miyamura2}.
In addition, one may note that the singular part $\theta^{\rm 
Mo}_{\mu}$ keeps essential contributions to
confining features of the Polyakov loop \cite{Suzuki} and finite 
quark condensate \cite{Miyamura2} in the finite temperature phase transition. 
Also, the SU(2) string tension is almost evaluated from the 
singular part in the MA gauge ($\sigma_{\rm Mo} \simeq 0.87 
\sigma_{\rm SU(2)}$) \cite{Bali}. 

In order to show the explicit contribution of monopoles for the
${\rm U_{A}}(1)$ anomaly,
we define two types of gauge configuration; the ``monopole removed 
({\it photon-dominating})'' 
link variable $U^{\rm Ph}_{\mu}$ and the ``photon removed
({\it monopole-dominating})'' 
link variable $U^{\rm Mo}_{\mu}$ as the corresponding SU(2) variables 
\cite{Miyamura1,Sasaki1}.
$U^{\rm Ph}_{\mu}$ is defined by removing the monopole 
contribution from the original gauge configuration as
%
%
\beqn
U^{\rm Ph}_{\mu}(n) \equiv U_{\mu}(n) \cdot u^{\dag\;\rm Mo }_{\mu}(n) 
\;\;.
\eeqn
On the other hand, $U^{\rm Mo}_{\mu}$ is defined by removing the 
photon contribution from the original gauge configuration as
%
%
\beqn
U^{\rm Mo}_{\mu}(n) \equiv U_{\mu}(n) \cdot u^{\dag\;\rm Ph }_{\mu}(n)\;\;.
\eeqn
Here, $u^{\;i}_{\mu}(n)\equiv \exp\{i\sigma_{3}\theta^{\;i}_{\mu}(n)\}$ 
($i=$ Ph or Mo).
It is noted that these definitions exactly correspond to the 
reconstruction of 
the resulting SU(2) variables 
from $u^{\;i}_{\mu}$ by multiplying the adjoint ``matter'' field
{\it in the Landau gauge} \cite{Sasaki1}:
%
%
\beqn
{\tilde U}^{\;i}_{\mu}(n) \equiv {\tilde M}_{\mu}(n)\exp\{i\sigma_{3}
\theta^{\;i}_{\mu}(n)\} \;\;,
\eeqn
where ${\tilde M}_{\mu}(n)=d(n)M_{\mu}(n)d^{\dag}(n)$  
with $d(n)=e^{i\varphi(n)\sigma_{3}}$
\footnote[2]{$\theta_{\mu}=\theta^{L}_{\mu}+\del_{\mu}\varphi
=\theta^{\rm Ph}_{\mu}+\theta^{\rm Mo}_{\mu}+\del_{\mu}\varphi$}
One can easily see the relation ; 
${\tilde U}^{\;i}_{\mu}(n)= d(n)U^{\;i}_{\mu}(n)d^{\dag}(n+\mu)$.
In this sense, we call $U^{\rm Mo}_{\mu}$ as the monopole-dominating gauge 
configuration (Mo part) and $U^{\rm Ph}_{\mu}$ as the photon-dominating gauge 
configuration (Ph part) respectively.
In previous publications, we found the corresponding topological 
charge, which can be classified by an ``integer'' value, in the background of 
$U^{\rm Mo}_{\mu}$
\cite{Miyamura1,Sasaki1}. 
On the other hand, the non-zero topological charge
was never found in the background of $U^{\rm Ph}_{\mu}$ 
\cite{Miyamura1,Sasaki1}

\section{Zero modes of the Dirac operator}

We study the low-lying eigenvalue spectra of the Dirac operator in 
the background of three types of configuration; 
the monopole-dominating gauge fields, 
the photon-dominating gauge fields and 
the original SU(2) gauge fields \cite{Sasaki1}.
For the Dirac operator on the lattice, 
we adopt the Wilson fermion 
\cite{Wilson}.
In the background of $U^{\;i}_{\mu}$ ($i=$ Ph or Mo)
and the original gauge fields; $U_{\mu}$, $D \kern -2.2mm {/}$
is expressed as
%
%
\beqn
D \kern -2.2mm {/}({n,\;m}) 
= \delta_{n,\;m}-\kappa \sum_{\mu} \left[
(1-\gamma_{\mu})U^{\;(i)}_{\mu}(n)\delta_{n+{\hat \mu},\;m} 
+(1+\gamma_{\mu})U^{\dag (i)}_{\mu}(n-{\hat \mu})
\delta_{n-{\hat \mu},\;m}\right] \;\;,
\eeqn
where $\kappa$ is the hopping parameter.
Although the Wilson fermion does not have the chiral symmetry in
the naive argument, the partial symmetry restoration would
be realized near the critical value $\kappa_{c}$ where 
the pseudo-scalar mass vanishes 
\cite{Kawamoto}.

The operator $D \kern -2.2mm {/}$ loses a feature as the hermitian operator 
owing to the discretization of the space-time. 
However, one can easily find that $\gamma_{5}{D \kern -2.2mm 
{/}}^{\dag}\gamma_{5}= {D \kern -2.2mm {/}}$.
We then examine the eigenvalue spectrum of the hermitian operator
$\gamma_{5}{D \kern -2.2mm 
{/}}$ by using the Lanczos algorithm.
We can identify the fermionic zero-modes in 
the following procedure.
First, the existence of zero modes 
could be found by the zero-line crossing in eigenvalue spectra
of $\gamma_{5}{D \kern -2.2mm {/}}$ through
the variation of the hopping parameter 
around $\kappa_{c}$ 
\cite{Itoh}.
Then, the chirality of zero-modes could be defined by the slope of 
the eigenvalue spectrum 
\cite{Itoh}.

We generate the gauge configuration by using the Wilson action on an
$8^4$ lattice with $\beta = 2.4$. 
As the hopping parameter, we change
within the range; $0.130 \leq \kappa \leq 0.180$.
(In ref.\cite{Fukugita}, $\kappa_{c}=0.175 \pm 0.002$ has been 
obtained at $\beta=2.4$ on a $5^3 \times 10$ lattice.)
We measure the eigenvalue of the 
operator $\gamma_{5}{D \kern -2.2mm {/}}$ 
in the background of $U^{\;i}_{\mu}$ ($i=$ Ph or Mo) and also 
in the original SU(2) gauge field; $U_{\mu}$ for 32 independent 
configurations {\it without any cooling} 
\cite{Sasaki1}.
Fig.\ref{fig:Zero_SU1}-\ref{fig:Zero_SU4} show the low-lying spectra in 4 independent 
gauge configurations as typical examples. In each configuration, we can
see (a) 1 zero mode of chirality $-$, (b) 2 zero modes of 
chirality $+$, (c) no zero mode and (d) 1 zero mode of chirality $+$. 
Shows that Fig.\ref{fig:Zero_Mo1}-\ref{fig:Zero_Mo4} and 
Fig.\ref{fig:Zero_Ph1}-\ref{fig:Zero_Ph4},
we can also observe the low-lying spectra 
in the background of the monopole-dominating gauge field and the 
photon-dominating gauge field, which are defined 
on the basis of the same gauge 
configurations in Fig.\ref{fig:Zero_SU1}-\ref{fig:Zero_SU4}. 
The same number of zero modes as be observed in 
Fig.\ref{fig:Zero_SU1}-\ref{fig:Zero_SU4} can be found 
in the background of $U_{\mu}^{\rm Mo}$ \cite{Sasaki1}.
This remarkable coincidence for the number of zero modes and its 
chirality is not well identified in 6 configurations, 
but is confirmed in all the rest 26 configurations \cite{Sasaki1}. 
However, we can not find the corresponding zero modes in the
background of $U_{\mu}^{\rm Ph}$ within 32 configurations.
It is worth mentioning that this result is consistent with our previous 
works in Ref.\cite{Miyamura1,Sasaki1}, which showed 
that the non-zero value of 
the topological charge was never found in any photon-dominating
gauge configuration after several cooling sweeps.
Thus, we can interpret that instantons can not 
live in the {\it monopole removed} gauge configuration.

\section{Summary}

We have investigated topological aspects of the QCD vacuum structure
by using the Monte Carlo simulation within the SU(2) gauge theory.
We defined two types of gauge configuration; the monopole removed 
gauge configuration and the photon removed gauge configuration after the 
MA gauge fixing.
We measured the fermionic zero modes
of the Dirac operator in each gauge field background 
without any cooling method.
In only the background of the monopole-dominating gauge field, 
the explicit breaking of the ${\rm U}_{\rm A}(1)$ symmetry
occurs owing to the existence of the fermionic zero modes. 
On the other hand, we can never find the corresponding zero modes in 
the background where monopole contributions are completely removed.
These results imply that topological features are inherited in the
monopole-dominating (photon removed) gauge field, but spoiled in the 
photon-dominating (monopole removed) gauge field since the 
${\rm U}_{\rm A}(1)$ anomaly is related to
topological objects, \ie instantons. Of course, we must not forget 
that an $8^4$ lattice at $\beta=2.4$ in our simulation 
might be considerably small in physical units to produce definitive results.
However, it seems reasonable to suppose that our numerical data shows
the strong evidence for the topologically close relation 
between instantons and QCD-monopoles in the quantum vacuum of QCD 
after the typical abelian gauge fixing.
This statement is also strongly supported by our next study \cite{Sasaki2},
which shows that the topological charge can be approximately 
reconstructed from the monopole current and the abelian component of 
gauge fields in the MA gauge.

\section*{Acknowledgment}

We would like to acknowledge fruitful discussions with H. Suganuma
and H. Toki at Research Center for Nuclear Physics of Osaka
University, where all lattice QCD simulations in this paper have been 
performed on DEC Alpha Server 8400 5/440.
One of the authors (S.S.) had been supported by Research Fellowships of the
Japan Society for the Promotion of Science (JSPS) for Young Scientists
before changing his place of work from Yukawa Institute 
for Theoretical Physics (YITP) to RIKEN BNL Research Center.
S.S. would like to mention also the warm hospitality of
T. Matsui and K. Itakura during his residence at YITP in Kyoto University.
Another (O.M.) is supported by the Grant in Aid for Scientific
Research by the Ministry Education (A) (No.0830424).

%

\newpage
\centerline{\large FIGURE CAPTIONS}
\begin{description}

\vspace{1.0cm}
\item[FIG.1.]
\begin{minipage}[t]{13cm}
\baselineskip=20pt
Typical examples of low-lying spectra of $\gamma_{5} D \kern -2.2mm {/}$
through the variation of $\kappa$
in the background of the original SU(2) gauge field on an $8^4$ lattice at 
$\beta=2.4$.
\end{minipage}

\vspace{1.0cm}
\item[FIG.2.]
\begin{minipage}[t]{13cm}
\baselineskip=20pt
Low-lying spectra of $\gamma_{5} D \kern -2.2mm {/}$
through the variation of $\kappa$
in the background of $U^{\rm Mo}_{\mu}$ corresponding to the
examples of (a)-(d) in Fig.1.
\end{minipage}

\vspace{1.0cm}
\item[FIG.3.]
\begin{minipage}[t]{13cm}
\baselineskip=20pt
Low-lying spectra of $\gamma_{5} D \kern -2.2mm {/}$
through the variation of $\kappa$
in the background of $U^{\rm Ph}_{\mu}$ corresponding to the
examples of (a)-(d) in Fig.1.
\end{minipage}
\end{description}

\newpage

\centerline{\Large FIG.1 (Phys.Lett.B) Shoichi Sasaki \etal}

\vspace*{2.5cm}
{\setcounter{enumi}{\value{figure}}
\addtocounter{enumi}{1}
\setcounter{figure}{0}
\renewcommand{\thefigure}{\arabic{enumi}(\alph{figure})}

%
%
\noindent
\begin{minipage}{\minitwocolumn}
\centerline{\epsfxsize=7.5cm
\epsfbox{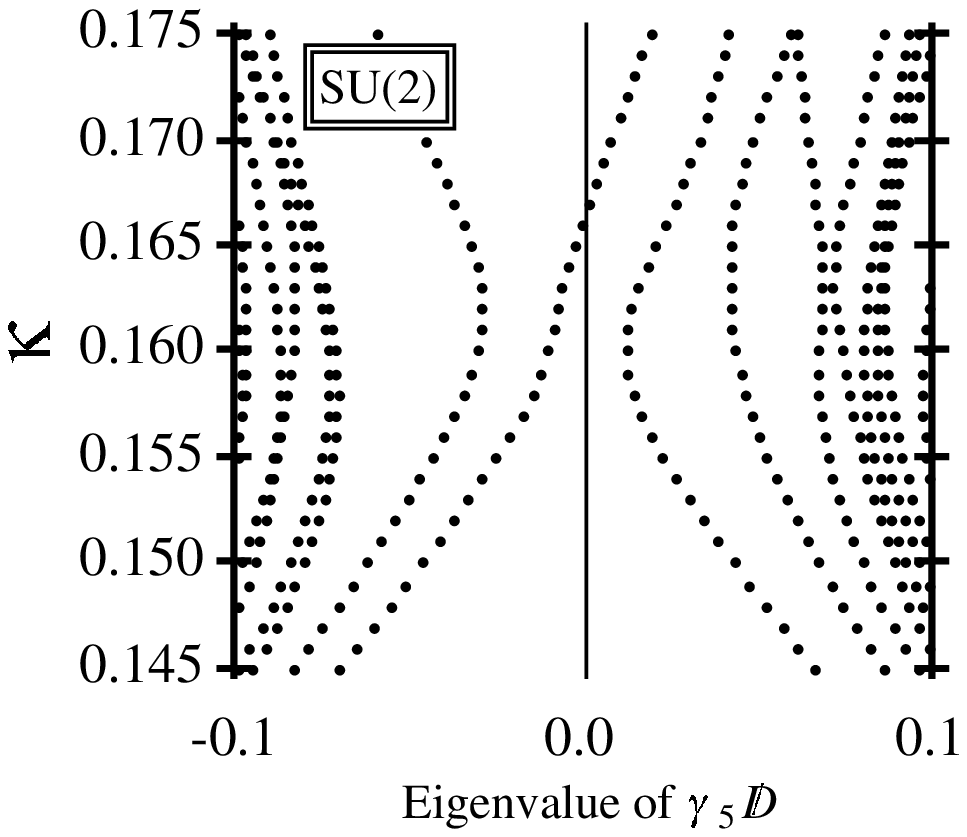}}
\vspace{0.5cm}
\centerline{\fcaption{\label{fig:Zero_SU1}}}
\end{minipage}
\hspace{\columnsep}
\begin{minipage}{\minitwocolumn}
\centerline{\epsfxsize=7.5cm
\epsfbox{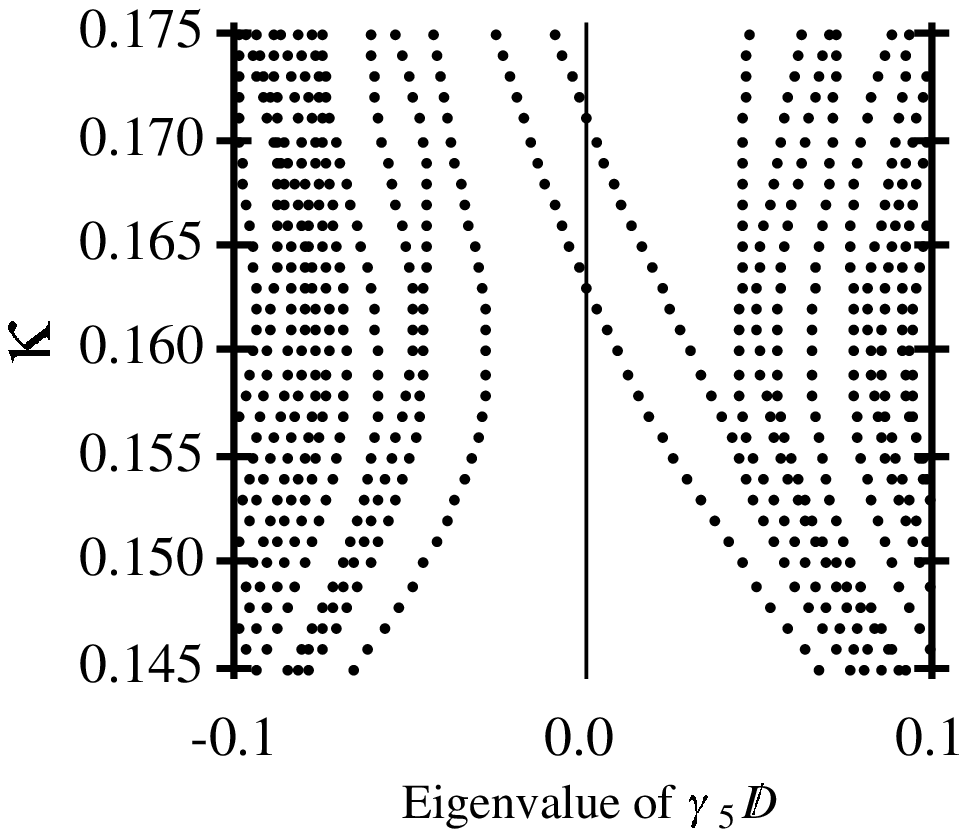}}
\vspace{0.5cm}
\centerline{\fcaption{\label{fig:Zero_SU2}}}
\end{minipage} 

\vspace*{2.5cm}
\noindent
%
%
\begin{minipage}{\minitwocolumn}
\centerline{\epsfxsize=7.5cm
\epsfbox{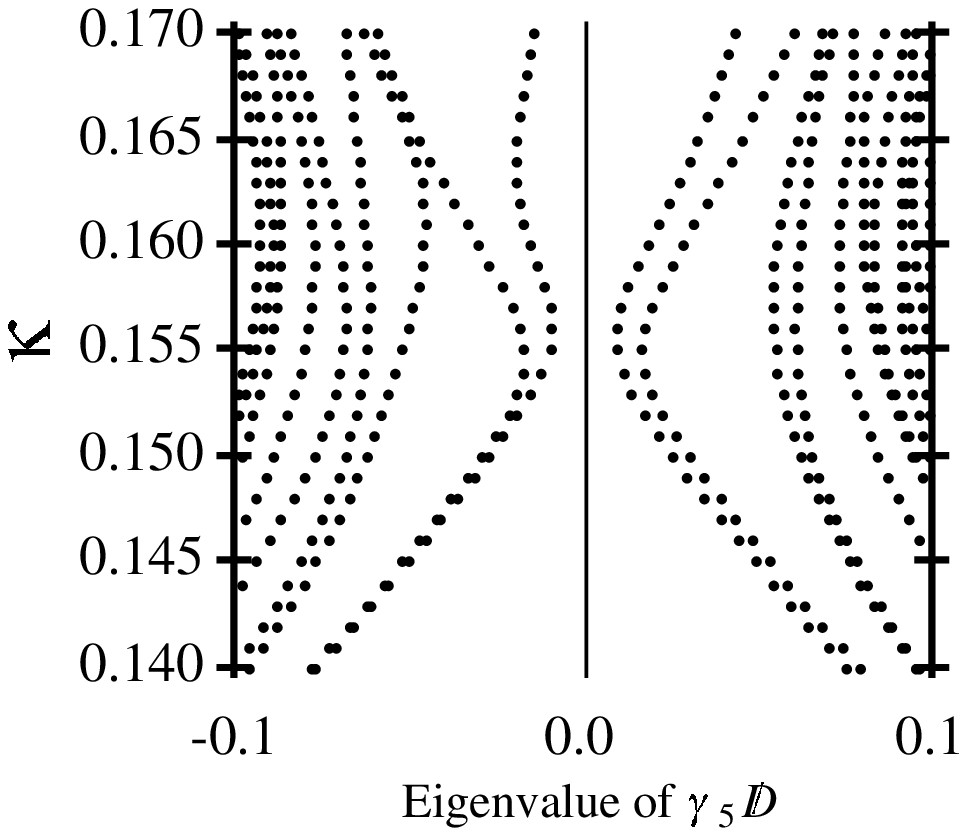}}
\vspace{0.5cm}
\centerline{\fcaption{\label{fig:Zero_SU3}}}
\end{minipage}
\hspace{\columnsep}
\begin{minipage}{\minitwocolumn}
\centerline{\epsfxsize=7.5cm
\epsfbox{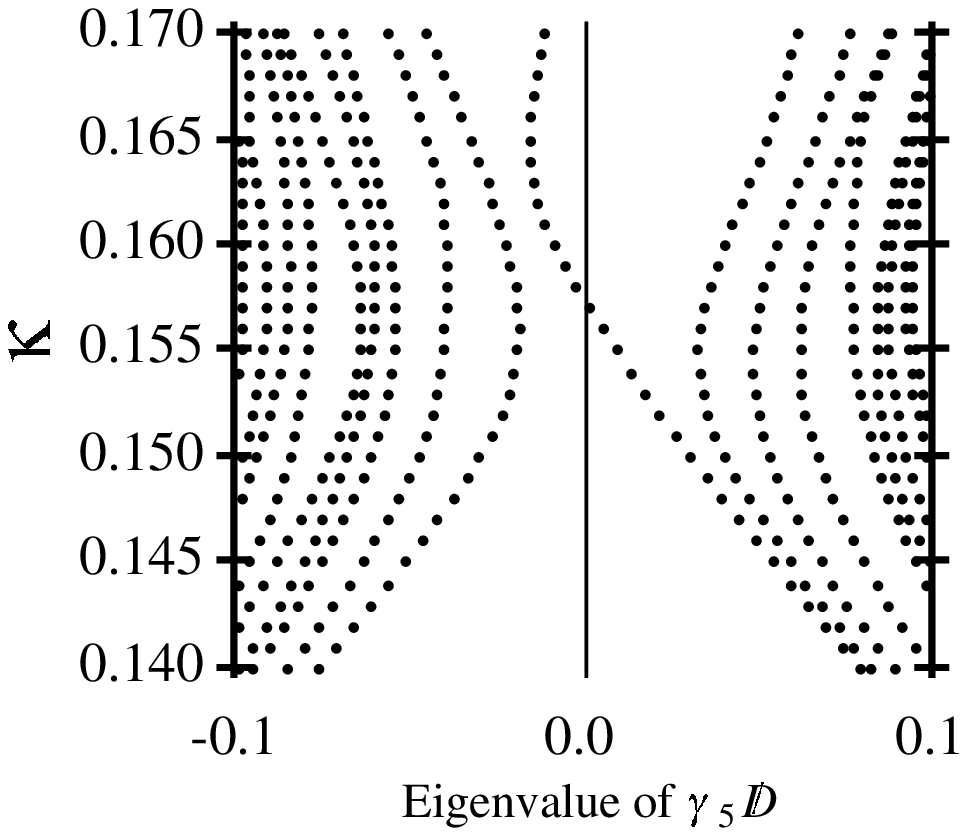}}
\vspace{0.5cm}
\centerline{\fcaption{\label{fig:Zero_SU4}}}
\end{minipage} 
\setcounter{figure}{\value{enumi}}
}
\newpage

\centerline{\Large FIG.2 (Phys.Lett.B) Shoichi Sasaki \etal}

\vspace*{2.5cm}
{\setcounter{enumi}{\value{figure}}
\addtocounter{enumi}{1}
\setcounter{figure}{0}
\renewcommand{\thefigure}{\arabic{enumi}(\alph{figure})}

%
%
\noindent
\begin{minipage}{\minitwocolumn}
\centerline{\epsfxsize=7.5cm
\epsfbox{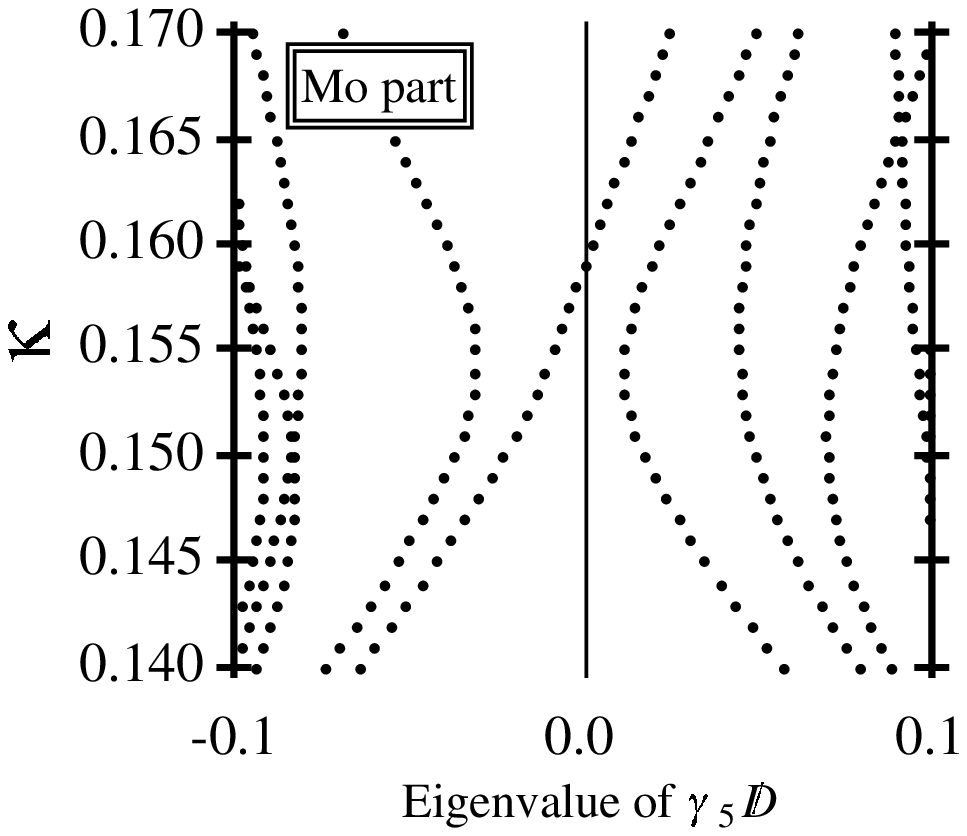}}
\vspace{0.5cm}
\centerline{\fcaption{\label{fig:Zero_Mo1}}}
\end{minipage}
\hspace{\columnsep}
\begin{minipage}{\minitwocolumn}
\centerline{\epsfxsize=7.5cm
\epsfbox{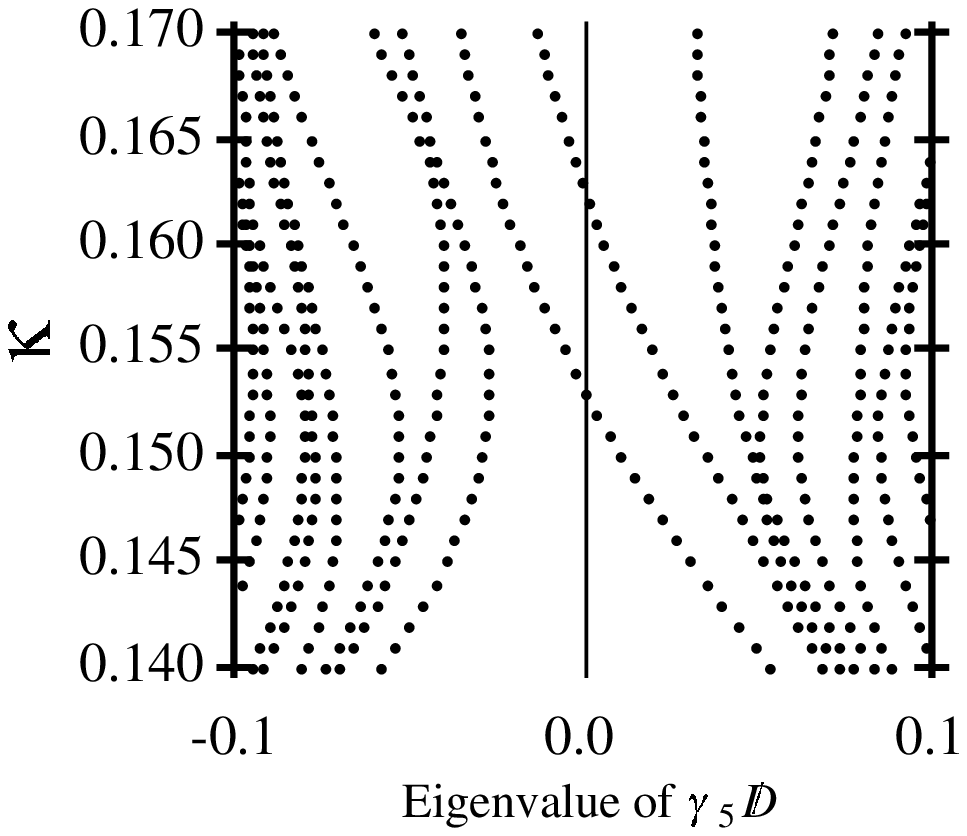}}
\vspace{0.5cm}
\centerline{\fcaption{\label{fig:Zero_Mo2}}}
\end{minipage}

\vspace*{2.5cm}
\noindent
%
%
\begin{minipage}{\minitwocolumn}
\centerline{\epsfxsize=7.5cm
\epsfbox{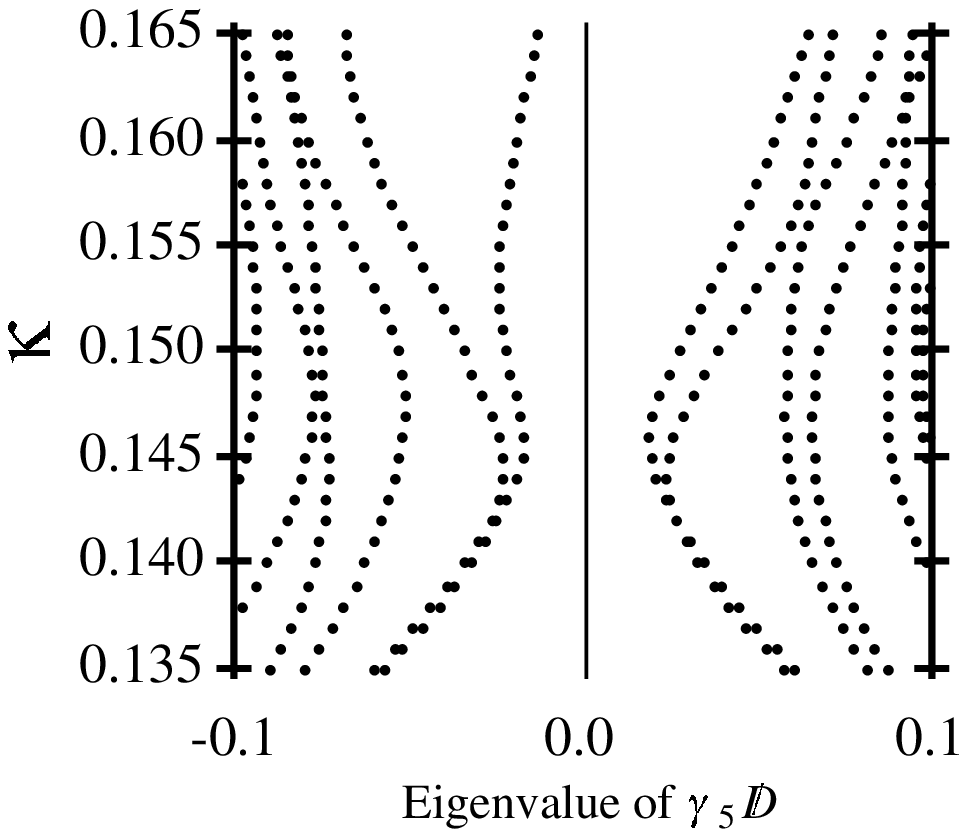}}
\vspace{0.5cm}
\centerline{\fcaption{\label{fig:Zero_Mo3}}}
\end{minipage}
\hspace{\columnsep}
\begin{minipage}{\minitwocolumn}
\centerline{\epsfxsize=7.5cm
\epsfbox{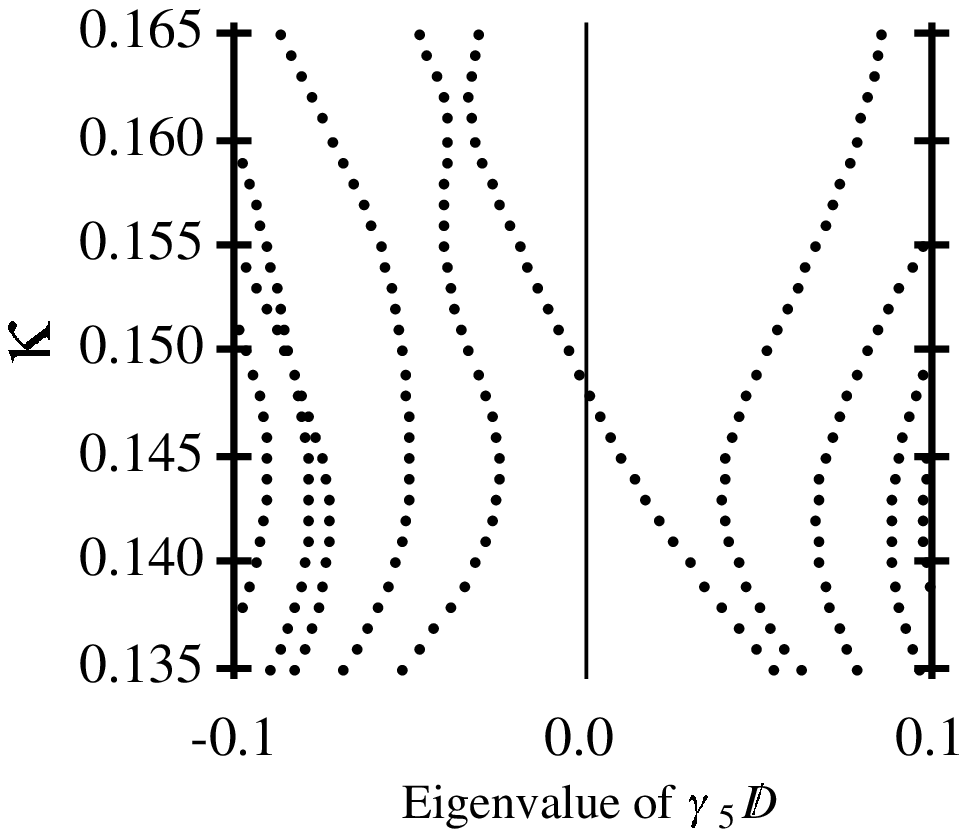}}
\vspace{0.5cm}
\centerline{\fcaption{\label{fig:Zero_Mo4}}}
\end{minipage}
\setcounter{figure}{\value{enumi}}
}
\newpage

\centerline{\Large FIG.3 (Phys.Lett.B) Shoichi Sasaki \etal}

\vspace*{2.5cm}
{\setcounter{enumi}{\value{figure}}
\addtocounter{enumi}{1}
\setcounter{figure}{0}
\renewcommand{\thefigure}{\arabic{enumi}(\alph{figure})}

%
%
\noindent
\begin{minipage}{\minitwocolumn}
\centerline{\epsfxsize=7.5cm
\epsfbox{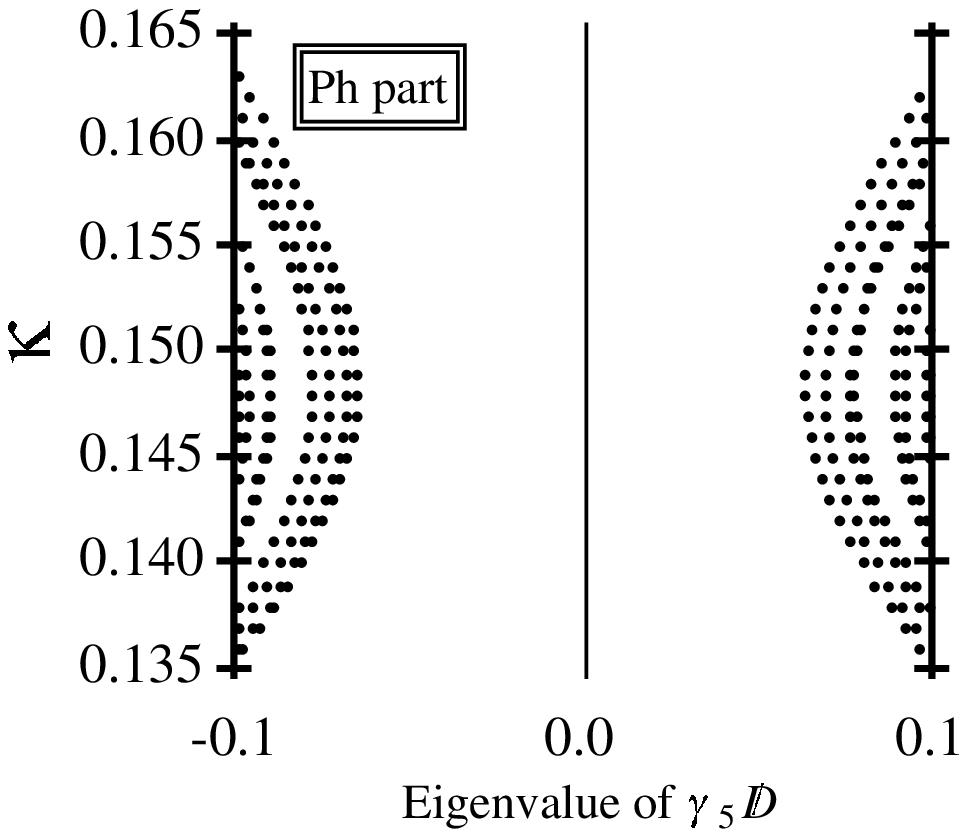}}
\vspace{0.5cm}
\centerline{\fcaption{\label{fig:Zero_Ph1}}}
\end{minipage}
\hspace{\columnsep}
\begin{minipage}{\minitwocolumn}
\centerline{\epsfxsize=7.5cm
\epsfbox{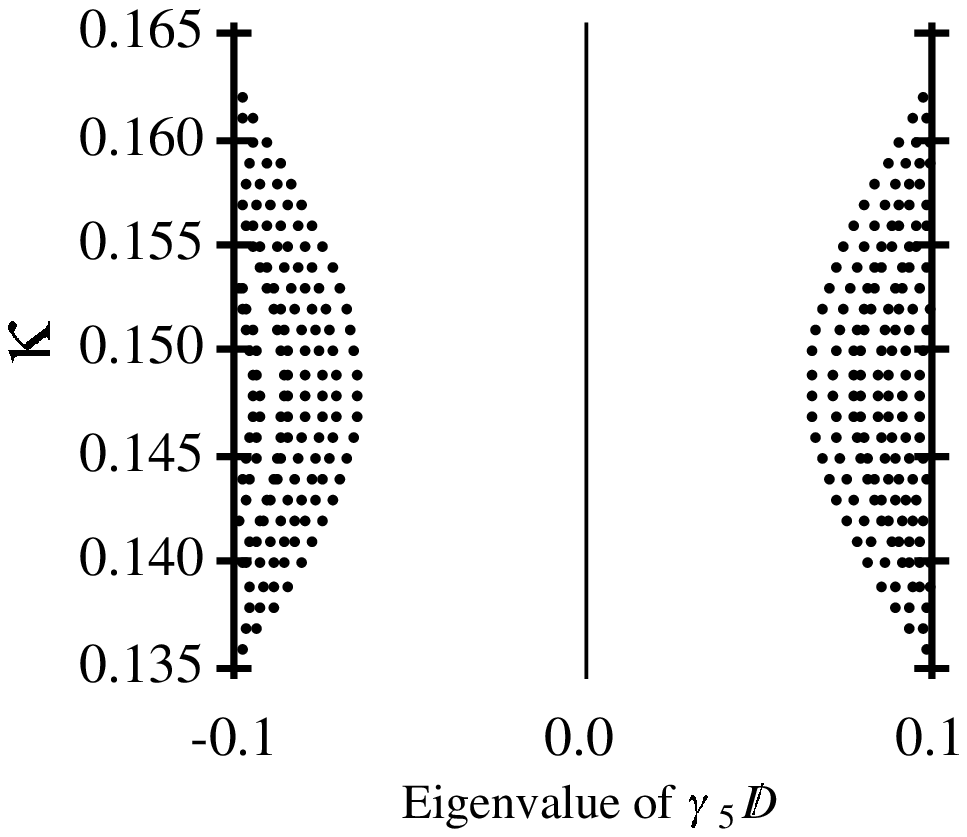}}
\vspace{0.5cm}
\centerline{\fcaption{\label{fig:Zero_Ph2}}}
\end{minipage} 

\vspace*{2.5cm}
\noindent
%
%
\begin{minipage}{\minitwocolumn}
\centerline{\epsfxsize=7.5cm
\epsfbox{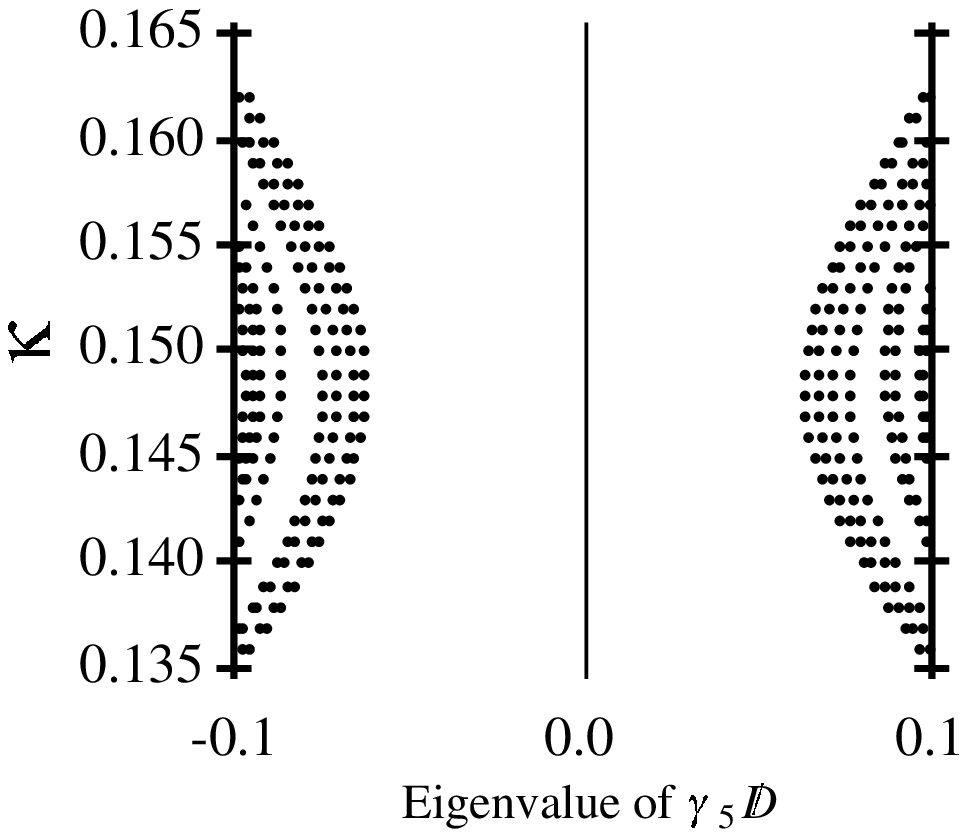}}
\vspace{0.5cm}
\centerline{\fcaption{\label{fig:Zero_Ph3}}}
\end{minipage}
\hspace{\columnsep}
\begin{minipage}{\minitwocolumn}
\centerline{\epsfxsize=7.5cm
\epsfbox{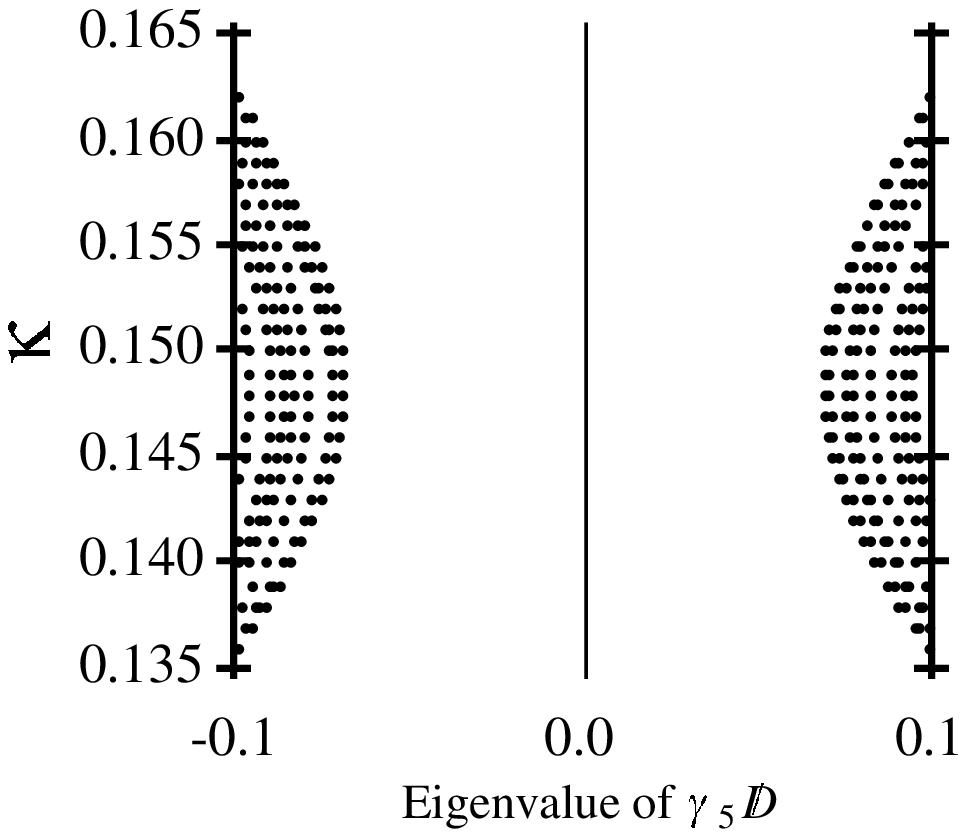}}
\vspace{0.5cm}
\centerline{\fcaption{\label{fig:Zero_Ph4}}}
\end{minipage} 
\setcounter{figure}{\value{enumi}}
}
\end{document}